\documentclass[aps,prl,nofootinbib,preprintnumbers,amsmath,amssymb,latexsym,array,enumerate,letter,twocolumn,superscriptaddress]{revtex4}
\usepackage{amssymb}
\usepackage{amsmath}
\usepackage{epsfig}
\usepackage{hyperref}
\usepackage{breakurl}
\usepackage{color}

\makeatletter
\def\simgt{\mathrel{\lower2.5pt\vbox{\lineskip=0pt\baselineskip=0pt
           \hbox{$>$}\hbox{$\sim$}}}}
\def\simlt{\mathrel{\lower2.5pt\vbox{\lineskip=0pt\baselineskip=0pt
           \hbox{$<$}\hbox{$\sim$}}}}
\makeatother

\newcommand{\be}{\begin{equation}}
\newcommand{\ee}{\end{equation}}
\newcommand{\bea}{\begin{eqnarray}}
\newcommand{\eea}{\end{eqnarray}}
\newcommand{\beq}{\begin{eqnarray}}
\newcommand{\eeq}{\end{eqnarray}}

\def\lsim{\mathrel{\rlap{\lower4pt\hbox{\hskip1pt$\sim$}}
     \raise1pt\hbox{$<$}}}         
\def\gsim{\mathrel{\rlap{\lower4pt\hbox{\hskip1pt$\sim$}}
     \raise1pt\hbox{$>$}}}         

\begin{document}

\title{Detecting Axion-like Dark Matter with Linearly Polarized Pulsar Light}

\author{Tao Liu}
\email{taoliu@ust.hk}
\affiliation{Department of Physics, The Hong Kong University of Science and Technology, Hong Kong S.A.R., P.R.China}

\author{George Smoot}
\email{gfsmoot@ust.hk}
\affiliation{Institute for Advanced Study, The Hong Kong University of Science and Technology, Hong Kong S.A.R, P.R.China }
\affiliation{Department of Physics, The Hong Kong University of Science and Technology, Hong Kong S.A.R., P.R.China}

\author{Yue Zhao}
\email{zhaoyue@physics.utah.edu}
\affiliation{Department of Physics and Astronomy, University of Utah, Salt Lake City, UT 84112, USA}

\begin{abstract}
Non-relativistic QCD axions or axion-like particles are among the most popular candidates for cold Dark Matter (DM) in the universe. We proposed to detect axion-like DM, using linearly polarized pulsar light as a probe.  Because of birefringence effect potentially caused by an oscillating galactic axion DM background, when pulsar light travels across the galaxy, its linear polarization angle may vary with time. With a soliton+NFW galactic DM density profile, we show that this strategy can potentially probe an axion-photon coupling as small as $\sim 10^{-13}$ GeV$^{-1}$ for axion mass $m_a \sim 10^{-22}-10^{-20}$ eV, given the current measurement accuracy. 
An exclusion limit stronger than CAST ($ \sim 10^{-10}$ GeV$^{-1}$) and SN1987A ($ \sim 10^{-11}$ GeV$^{-1}$) could be extended up to $m_a \sim 10^{-18}$ eV and $\sim 10^{-19}$ eV, respectively. 

\end{abstract}

\maketitle

\section{Introduction}

Non-relativistic QCD axions or axion-like particles (below we will not distinguish these two concepts for simplicity) have been known for decades to be able to serve as a candidate of cold Dark Matter (DM)~\cite{Preskill:1982cy,Abbott:1982af,Dine:1982ah}. In this context, the axion stability at cosmological time scale is protected by its large decay constant and tiny mass, whereas its non-relativistic properties may result from specific production mechanisms such as misalignment~\cite{Preskill:1982cy,Abbott:1982af,Dine:1982ah}. As an initial condition of this mechanism, the population of axions starts as a coherent state. The evolution of such a state yields Cosmic Axion Background (CAB). Especially interesting, the small-scale structure problems in astrophysics (e.g., the ``cusp-core'' problem) which are challenging the weakly-interacting-massive-particle cold DM paradigm can be potentially addressed in a special scenario of axion DM, named ``fuzzy DM''~\cite{Hu:2000ke,Peebles:2000yy,Hui:2016ltb,Broadhurst:2018fei}. This requires the axions to be ultralight, with a mass $\sim 10^{-22}$ eV.  Then the formation of a cuspy galactic DM core is suppressed because of  quantum pressure. As recently revealed in a high-resolution cosmological simulation of the Schrodinger equation~\cite{Schive:2014dra}~\cite{Hui:2016ltb}, a cored solitonic DM density profile instead can be formed in the galactic center.  

The strategies to detect relic axions or the CAB are quite diverse, ranging from astrophysical observations, cosmological measurements to lab experiments (for reviews, see, e.g.,~\cite{Marsh:2015xka,Graham:2015ouw}). Most of them are based on the axion-photon interaction (for the detections which are based on the axion couplings with gluons and neutrons, see, e.g.,~\cite{Stadnik:2013raa,Abel:2017rtm}, and on the ones with gravity, see~\cite{Khmelnitsky:2013lxt,Porayko:2018sfa,DeMartino:2017qsa}), with the relevant Lagrangian given by
\begin{eqnarray}
L \sim -\frac{1}{4}F_{\mu\nu}F^{\mu\nu}+ \frac{1}{2}\partial^\mu
a\partial_\mu a - \frac{1}{2}m_a^2 a^2+\frac{g}{2}a F_{\mu\nu}\tilde
F^{\mu\nu}  
\label{Pinteraction}
\end{eqnarray}
Here $a$ is the axion field, $m_a$ is its mass parameter, $F_{\mu\nu}$is
the electromagnetic (EM) field strength, and $\tilde F_{\mu\nu}$ is its dual. Searches based on this interaction can roughly fall into two categories. One is to convert the relic axions into EM signals in a laboratory or astrophysical magnetic field. The ADMX~\cite{Asztalos:2009yp} in operation is such a haloscope experiment~\cite{Sikivie:1983ip}. For ADMX, in order to enhance the conversion rate, the resonance frequency of the cavity needs to be tuned to match the axion oscillation frequency. This makes such experiment only accessible to a narrow range of axion mass. Similar axion-photon conversion process in the presence of external magnetic field has also been applied to detect non-relic axions, where the axions are generated either  in laboratory \cite{Bahre:2013ywa} or in astrophysical environment \cite{Arik:2013nya}. 

Another way is to measure the effect of cosmological birefringence. When light travels in the CAB, its left- and right-handed circular polarization modes will receive opposite corrections due to their dispersion relations.  Hence, if the light is linearly polarized, its polarization angle will be shifted~\cite{Carroll:1989vb,Carroll:1991zs,Harari:1992ea}. Such a birefringence effect has been extensively applied to detect the CAB, using the $B$-mode polarization of Cosmic Microwave Background (CMB), the radio/ultraviolet polarization of radio galaxy and active galactic nucleus (AGN) (for a review, see, e.g.,~\cite{Marsh:2015xka,Alighieri:2010pu,Galaverni:2018zcm}), and others~\cite{Plascencia:2017kca,Obata:2018vvr,Fujita:2018zaj,Liu:2018icu}. 

In this letter, we instead propose to detect the relic axions utilizing the birefringence effect on linearly polarized pulsar light. Because of low-scale Peccei-Quinn symmetry breaking or DM clustering, the CAB could be spatially inhomogeneous. But this effect is usually neglected in the CMB detection (see, e.g.,~\cite{Marsh:2015xka,Finelli:2008jv,Sigl:2018fba,Pospelov:2008gg}) to avoid analysis complexity. The probe of linearly polarized pulsar light however may allow us to address this poperly, given the  knowledge on galactic DM energy density profile. This probe also benefits the CAB detection in several other aspects. Both second pulsars (SPs) and millisecond pulsars (MPs) are known to be stable astrophysical sources of linearly polarized light. Their repeating light pulses potentially enable us to measure time variation of the linear polarization angle which could be induced by the CAB oscillation. Additionally, more than two thousands of SPs and MPs have been discovered so far in our galaxy, and many more are expected to be explored in the near future~\cite{Beck:2009ew}, e.g., by FAST~\cite{fast}. The richness of such light sources allows us to improve detection sensitivities, by correlating the observations to suppress both astrophysical background and instrumental uncertainties.

\section{Cosmological Birefringence}

While traveling through an oscillating CAB, the light with different circular polarizations receives opposite corrections to its dispersion relation. At leading order, its dispersion relation is given by 
\begin{eqnarray}\label{dispersion}
\omega \simeq k\pm g(\frac{\partial a}{\partial t}+\vec\nabla
a\cdot \frac{\vec k}{k}) \ . 
\end{eqnarray}
The axion DM is highly non-relativistic, characterized by the virial velocity of galaxies. In our galaxy, the virial velocity is $v_{vir}\simeq 230 \textrm{km}/\textrm{s}\sim O(10^{-3})c$ (below we will neglect the potential difference of this value in galactic central region). So we can safely neglect the last term in Eq. (\ref{dispersion}). 
If the light is linearly polarized, these corrections will result in a birefringence effect, say, a rotation of its polarization angle. This effect is independent of light frequency, since $\frac{\partial a}{\partial t}$ is just a description of the CAB time variation.   

To quantitatively calculate this rotation, we need to model the CAB within the galaxy. Locally, the non-relativistic CAB can be parametrized as a planewave 
\begin{eqnarray}\label{axionBG}
a(x,t)\simeq a_{0} (x) \cos(m_a t+\theta) \ ,
\end{eqnarray}
with the phase $\theta$ being an approximate constant. Here we have neglected the characteristic momentum and kinetic energy of the CAB. This approximation breaks down when spatial separation is greater than coherence length ($l_c$) or temporal separation is greater than coherence time ($t_c$). Here $l_c$ and $t_c$ are determined by virial momentum and kinetic energy of the axion DM, i.e., $l_c = \frac{2\pi}{k_a}$ and $t_c = \frac{4\pi m_a}{k_a^2}$, respectively.  They together define the CAB coherence region. The phase values of the plane wave are uncorrelated in different coherence regions. 

Analytically, this rotation is described by a time integral of $\frac{\partial a}{\partial t}$ over the traveling of light from its source to the destination~\cite{Carroll:1989vb,Carroll:1991zs,Harari:1992ea,Marsh:2015xka}, up to a constant factor. If the CAB is continuously differentiable w.r.t time, we have 
\begin{eqnarray}\label{ChangeOneCoh}
\Delta\phi  \simeq g\int_{t_f}^{t_i} \frac{\partial}{\partial t}
a(x,t) dt =  g [a(x_f, t_f)-a(x_i, t_i)].
\end{eqnarray}
$\Delta\phi $ depends on the CAB profile at $(x_i, t_{i})$ and $(x_f, t_{f})$, that is, the position and moment that the light is emitted and observed.  For non-relativistic axion DM, $a_0(x)$ can be related to local DM energy density $\rho(x)$ via an approximate relation $\rho (x) \approx \frac{1}{2}m_a^2 a_0 (x)^2$. If the DM energy density at the observation point is much smaller than that at the emission point, we have 
\begin{eqnarray}\label{ChangeOneCoh1}
\Delta\phi  &\simeq&- g \frac{\sqrt{2\rho_i}}{m_a} \cos(m_a t_i+\theta_i).
\end{eqnarray}
As above, the subscript ``$i$'' represents that the relevant quantities are defined at the initial moment and position of light. 
Instead, if the axion DM energy densities are comparable at the initial and final points of light, we have 
\begin{eqnarray}\label{ChangeOneCoh2}
\Delta\phi  \simeq g \frac{\sqrt{2\rho_i}}{m_a} [\cos(m_a t_i+\theta_i)- \cos(m_a t_f+\theta_f)] \ .
\end{eqnarray}
Here $\theta_i$ and $\theta_f$ are uncorrelated if the light has traveled across multiple coherence regions along the line of sight before reaching us. 
Given the randomness of their values, we can use the standard deviation of $\Delta\phi$ to characterize its magnitude. This yields  
\begin{eqnarray}
\phi_{c} \equiv \sqrt{\langle \Delta\phi^2 \rangle} = g  \frac{\sqrt{ \rho_i +\rho_f}}{m_a} 
= \begin{cases}  g \frac{\sqrt{ \rho_i}}{m_a} \ , & \rho_i \gg \rho_f \ . \\ 
g  \frac{\sqrt{2 \rho_i}}{m_a} \ , & \rho_i = \rho_f \ .
\end{cases}
\end{eqnarray}

If the initial polarization angle of the light were known, one would have been able to probe the CAB by comparing it with the observed value. However, this  information is usually unavailable for astrophysical sources such as pulsars. Thus we propose to detect cosmological birefringence by measuring its time variation\footnote{Similar idea was shared by a recent study using the AGN light as the probe~\cite{Ivanov:2018byi}.}.  

If the temporal separation between two sequential light signals is much smaller than the coherence time, i.e. $\Delta t \ll t_c$, the rotations of their polarization angles are correlated, yielding 
\begin{eqnarray}\label{CorrPolar}
\Delta \Phi 
 =  \Delta\phi_{2}-\Delta\phi_{1} =  \Delta\Phi_0 \sin\bigg(\frac{m_a\Delta t}{2}\bigg)  \ .
 \end{eqnarray}
Here 
\begin{eqnarray} 
\Delta\Phi_0 &=& - \frac{2\sqrt{2}g}{m_a} \Big( \sqrt{\rho_f} \sin(m_a t_{f,+} +\theta_f)  
\nonumber \\ && - \sqrt{\rho_i}  \sin(m_a t_{i,+} +\theta_i)  \Big ) 
\end{eqnarray}
is the magnitude of such time variation, with $t_{f,+} =\frac{t_{f,1}  + t_{f,2}}{2}$ and $t_{i,+} =\frac{t_{i,1}  + t_{i,2}}{2}$. Similar to $\Delta \phi$, this quantity can be characterized by its standard deviation over all possible values of $\theta_i$ and $\theta_f$, given by
\begin{eqnarray} \label{equal} 
\Phi_c \equiv \sqrt{\langle\Delta\Phi_0^2\rangle} = \frac{2 g}{m_a} \sqrt {\rho_i + \rho_f} = 2 \phi_c \ .
\end{eqnarray}

The time-varying effect in this observable is described by $\sin\big(\frac{m_a\Delta t}{2}\big)$, as indicated in Eq. (\ref{CorrPolar}). This sinusoidal factor could be spoiled if the two sequential light signals receive sizable uncorrelated corrections during their whole journey to the Earth. This may happen since these two light signals do not experience exactly the same CAB profile due to its evolution. But, recall 
\begin{eqnarray} 
\Delta\Phi & = & g [a(x_{f}, t_{f,2})-a(x_{f}, t_{f,1})] \nonumber\\
&&- g [a(x_{i}, t_{i,2})-a(x_{i}, t_{i,1})] \ ,
\end{eqnarray}
is determined by the temporal and spatial variations of the CAB at the emission and observation points only. As long as $\Delta t$ is much smaller than the coherence time, the correlation loss caused by the CAB evolution can be neglected. Instead, a stronger limitation for the application of this observable could arise from the requirement of time resolution for the probe.  To pursue the measurement, one needs $\Delta t$, the time resolution of the probe, to be much smaller than $\frac{4\pi}{m_a}$, the characteristic time of this observable. As we will discuss below, this sets up an upper limit for $m_a$, that is, 
\begin{eqnarray}
m_a \ll \frac{4\pi}{\Delta t} \ , \label{limit}
\end{eqnarray}
where this observable can be applied.

\section{Pulsar-based Detection}

Pulsars are one of the main astrophysical sources of linearly polarized light. In order to obtain sensible information on the properties of pulsar light, e.g. the degree of its linear polarization and the relevant polarization angle, we usually measure them by first combining hundreds of successive pulses into one bin and then taking the average over the pulses in each bin. Such a procedure yields a time interval $\Delta t \sim \mathcal O(100)$s between two adjacent bins for SPs, and $\Delta t \sim \mathcal O(0.1)$s for MPs. For concreteness, we define the information provided by each bin as one light signal, with $\Delta t = 100$s for SPs and $\Delta t = 0.1$s for MPs. This sets up an upper limit $m_a \ll 8.3 \times 10^{-17}$ eV for SPs and $m_a \ll 8.3 \times 10^{-14}$ eV for MPs, bases on Eq. (\ref{limit}) where one can apply the proposed strategy. 

Currently the accuracy of measuring the linear polarization angle of pulsar light is $\sim 1^\circ$ or $\sim 0.017$ rad (see, e.g.~\cite{Moran:2013cla}). 
In this study, we consider two benchmark pulsars both of which sit along the line of sight to the galactic center. 
The relevant information about these two pulsars is provided in Table~\ref{tab:benchmarks}. 
Recall, the pulsar closest to the galactic center could be only $\sim 1$ pc away from Sgr $A^*$~\cite{Rajwade:2016cto};  
and the ones closest to the Earth are $\sim \mathcal O(100)$ pc far (e.g., PSR J0108-1431 is at a distance $\sim 130$ pc to the Earth~\cite{Posselt:2008ka}). The two benchmark pulsars thus represent a broad class of pulsars known to us.

\begin{table} 
\resizebox{0.6 \columnwidth}{!}{%
\begin{tabular}{|c|c|c|c|}
\hline
 & $R_j$ (pc) & $d_j$ (pc) & $m_{a, j}$ (eV)
\\ \hline \hline
$P_1$ & 1 & 8000 & $5.3\times 10^{-20}$ 
\\ \hline
$P_2$ & 7000 & 1000 &  $7.5\times 10^{-24}$ 
 \\   \hline  
\end{tabular}
}
\caption{Benchmark pulsars, denoted as $P_j$, with $j=1, 2$. $R_j$ is the distance of the benchmark pulsars to the galactic center. $d_j$ is their distance to the Earth. $m_{a,j}$ is the axion mass yielding $l_c = R_j$.} 
\label{tab:benchmarks}
\end{table}

For regions far away from the galaxy center, the DM density distribution can be approximately described by the NWF profile. 
A cored solitonic profile may take over at $r< l_c$, for ultralight axion DM,  due to quantum pressure~\cite{Schive:2014dra}. 
Given that an exact description is still absent in literatures, we simply parametrize the DM density distribution with a flat solitonic 
profile for $r < l_c$~\cite{Schive:2014dra} and an NFW profile for $r > l_c$, $i.e.$, (also see~\cite{Marsh:2015xka}) 
\begin{equation}
  \rho(x)=\begin{cases}
    0.019(\frac{m_a}{m_{a,0}})^{-2}(\frac{l_c}{{\rm 1 kpc}})^{-4}M_{\odot} \textrm{pc}^{-3}, & \text{for $r < l_c$}.\\
    \frac{\rho_0}{r/R_H(1+r/R_H)^2}, & \text{for $r > l_c$} \ .
  \end{cases}
\end{equation} 
Here $m_{a,0} = 10^{-22}$eV is a reference value for axion mass. 
$\rho_0=1.4\times 10^7 M_\odot/\textrm{kpc}^3$ and $R_H=16.1\textrm{kpc}$ are assumed for the NFW profile in our galaxy~\cite{Nesti:2013uwa}. 
Though a smooth transition between the soliton and NFW profiles should exist in a realistic case, we will tolerate this inaccuracy in this study, considering 
that the observable $\Phi_c$ defined in Eq. (\ref{equal}) only depends on physics at the initial and final traveling points of pulsar light.  

The total change of the linear polarization angle of pulsar light during its traveling to the Earth is then 
\begin{equation} \label{eq:or1}
\Phi_c^j =  \begin{cases}
11 \ \textrm{rad}\bigg(\frac{g}{g_{\rm CAST}}\bigg) ,  & \text{$m_a < m_{a,j}$} \ .  \\
2.7 \ \textrm{rad}\bigg(\frac{g}{g_{\rm CAST}}\bigg) \left( \frac{m_{a,0}}{m_a}\right) \\ \times \left( \frac{R_H}{R_j}  + \frac{R_H}{R_{e}}  \right)^{1/2}, & \text{$m_a > m_{a,j}$} 
 \ .  \end{cases}  
\end{equation} 
Here $j$ labels the benchmark pulsars. $g_{\rm CAST}=6.6\times 10^{-11} {\rm \ GeV}^{-1}$ is the state-of-the-art CAST limit for $m_a < 0.02$ eV at 95\% C.L.~\cite{Anastassopoulos:2017ftl}. $R_e\approx 8000$ pc is the distance between the Earth and the galactic center (see, e.g.,~\cite{Malkin:2013ac}). In deriving this formula, we have implicitly assumed $m_a \ge m_{a,0}$, the case that is interesting to us below. So, the case with $m_a < m_{a,j}$ is meaningful only if $m_{a,j} > m_{a,0}$ or the distance of the pulsar to the galactic center is smaller than 530 pc. In this context, the contribution to $\Phi_c^j$ which arises from the observation point at the Earth is negligibly small, and has been left out in Eq. (\ref{eq:or1}). $\Phi_c^j$ is thus independent of $m_a$. In the case with $m_a > m_{a,j}$, the pulsar and the Earth are in the NFW region. The contributions to $\Phi_c^j$ arising from both positions could be comparable if $R_j$ is not much smaller than $R_e$. So both of them are included in Eq. (\ref{eq:or1}), as denoted by the two terms in square root.

\section{Sensitivity Analysis}

\begin{figure}
   \includegraphics[width=1.0\columnwidth]{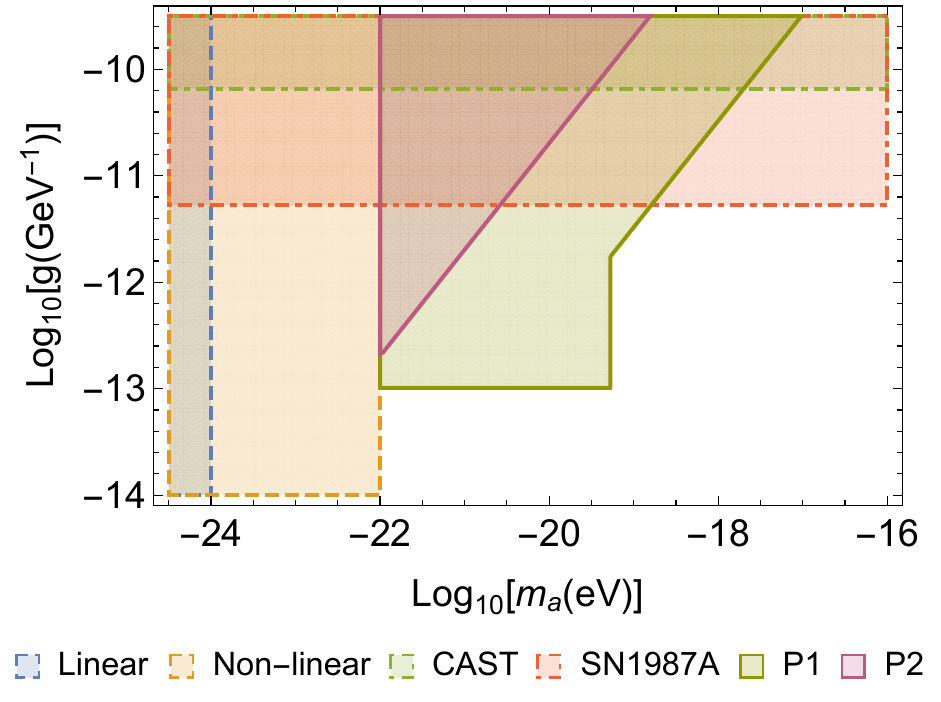}
  \caption{ Projected sensitivities to detect the CAB, using linearly polarized pulsar light as a probe, in the two benchmark scenarios: $P_1$ and $P_2$.
\label{fig:sen} }
\end{figure}

The projected sensitivities of the CAB detection using linearly polarized pulsar light, together with several constraints from astrophysical/cosmological observations, are shown in Fig. \ref{fig:sen}. 
The combination of the CMB and large-scale-structure observables in linear region, which measure the cosmic expansion rate and the structure growth, yields a constraint of $m_a > 10^{-24}$ eV~\cite{Hlozek:2014lca}. The constraints from non-linear clustering are even stronger, obtained by measuring the halo mass function. Based on  the structure suppression effect below the axion Jeans scale (similar to the free streaming effect of hot DM), the high-$z$ data has excluded the region with $m_a < 10^{-22}$ eV~\cite{Bozek:2014uqa,Schive:2015kza}. Note, here one has assumed that the axions compose all the DM. These exclusion limits are weakened by loosing this assumption. For example, if the axions contribute 50\% of the DM only, $m_a \sim 10^{-24}$ eV could be still allowed (see, e.g.,~\cite{Marsh:2015xka}). 

For comparison, the CAST limit~\cite{Anastassopoulos:2017ftl} and SN1987A constraint~\cite{Payez:2014xsa} are also included, with the latter being $g < 5.3\times 10^{-12}$ GeV$^{-1}$ for $m_a < 4.4 \times 10^{-10}$ eV~\cite{Payez:2014xsa}. Both limits are independent of the assumption of the axion DM, since the target axions are not from the relic, but sourced by astrophysical objects.

The exclusion limits set by the two benchmark pulsars $P_1$ and $P_2$ are shown, by comparing the characteristic quantity $\Phi_c^j$ with the current accuracy of measuring linear polarization angle of pulsar light.  
According to Eq. (\ref{CorrPolar}), the half period of its polarization-angle oscillation is $\sim 1.3$ yr, for $m_a = 10^{-22}$ eV. If $m_a < 10^{-22}$ eV, an observation period longer than $O(1)$ years is needed to measure this time-varying effect. So, we present the limits for $m_a \ge 10^{-22}$ eV only in Fig. \ref{fig:sen}. This is also consistent with that DM physics and the relevant observations favor more the parameter region with $m_a \sim$ and $> 10^{-22}$ eV~\cite{Hui:2016ltb,Marsh:2015xka}. 
The $P_1$ exclusion limit is universally stronger than the $P_2$ one. Its flat and slope parts result from the scenarios with the pulsar being positioned in the core soliton and NFW regions, respectively, with the threshold set by $m_{a,1} = 5.3\times 10^{-20}$ eV. 
In the flat region, $P_1$ sets its best limit $\sim 10^{-13}$ GeV for $g$. Compared to the constraints from CAST and SN1987A, it is improved by nearly three and two orders of magnitude, respectively. In the region above the threshold, the $P_1$ limit is quickly weakened by the $m_a^{-1}$ factor in Eq. (\ref{eq:or1}) as $m_a$ increases. As a comparison, the $P_2$ limit is set with the pulsar being positioned in the NFW region, due to $m_{a,2} < 10^{-22}$ eV. 
In this case the pulsar light does not pass a solitonic region any more. So the sensitivities are suppressed as $m_a$ increases.

\section{Summary and Outlook}

In this letter, we proposed to detect the axion DM, using linearly polarized pulsar light as a probe. Because of birefringence effect potentially caused by an oscillating galactic axion DM background, when pulsar light travels across the galaxy, its linear polarization angle may vary with time. With a soliton+NFW galactic DM density profile, we show that measuring the time variation of this polarization angle could probe an axion-photon coupling as small as $\sim 10^{-13}$ GeV$^{-1}$ for axion mass $m_a \sim 10^{-22} - 10^{-20}$ eV, given the current measurement accuracy. An exclusion limit stronger than CAST and SN1987A can be extended up to $m_a \sim 10^{-18}$ eV and $m_a \sim 10^{-19}$ eV, respectively. Note, these sensitivity limits will be linearly scaled as the accuracy of measuring the polarization angle of pulsar light improves in the future. 

Several issues are worthwhile to note. First, in this study, we treated the axion-photon coupling as a free parameter with a given $m_a$ value and neglected its potential influence for the axion relic abundance and hence for the galactic axion DM density profile. Also we ignored the subtleties raised due to axion self interaction~\cite{Sikivie:2009qn,Erken:2011dz,Guth:2014hsa}. Second, we didn't pursue a full exploration on the potential influence of astrophysical foreground and instrumental movement. For example, the polarization angle of the pulsar light can be changed by Faraday rotation effect, if the pulsar light travels across galactic magnetic fields with ionized gas. However Faraday rotation has a strong dependence on the light frequency and does not oscillate with time. The richness of the pulsars in our galaxy allows correlating the data analyses of multiple pulsars. The observation of the rotations with a universal oscillation frequency, within a broad frequency range, will be a strong signal for the CAB existence. Also, the potential error caused by instrumental drifting could be suppressed by correlating the observations of multiple pulsars. We leave a full study on these issues in a following-up work.

\begin{acknowledgments}
\subsection*{Acknowledgments} 
G.~Smoot is supported by IAS TT$\&$WF Chao Foundation. Y.~Zhao would like to thank the hospitality of the HKUST Jockey Club Institute for Advanced Study, where this project was initiated. 

\end{acknowledgments}

\end{document}